\documentclass[11pt,a4paper]{emulateapj}
\usepackage{natbib}
\usepackage{color}

\usepackage{epsfig}
\usepackage{amsmath}
\usepackage{amssymb}
\usepackage{natbib}
\usepackage{subfigure}
\usepackage{graphicx}
\def \taufif {$\tau^{15h^{-1}\rm{Mpc}}_{\rm{eff}}$}
\slugcomment{submitted to {\it The Astrophysical Journal}}

\newcommand{\civ}{C {\sc iv}}
\newcommand{\heii}{He {\sc ii}}
\newcommand{\ciii}{C {\sc iii}}
\newcommand{\oii}{O {\sc ii}}
\newcommand{\oiii}{O {\sc iii}}

\begin{document}

\def\mean#1{\left< #1 \right>}

\title{MApping the Most Massive Overdensities (MAMMOTH) II -- Discovery of an Extremely Massive Overdensity BOSS1441 at $z=2.32$.}
\author{Zheng Cai\altaffilmark{1,2,9}, Xiaohui Fan\altaffilmark{2}, Fuyan Bian\altaffilmark{3}, Ann Zabludoff\altaffilmark{2}, Yujin Yang\altaffilmark{4},  
  J. Xavier Prochaska\altaffilmark{1}, Ian McGreer\altaffilmark{2}, Zhen-Ya Zheng\altaffilmark{5,6}, Nobunari Kashikawa\altaffilmark{7}, Ran Wang\altaffilmark{8}, Brenda Frye\altaffilmark{2}, Richard Green\altaffilmark{2},  Linhua Jiang\altaffilmark{8}}
\affil{$^1$ UCO/Lick Observatory, University of California, 1156 High Street, Santa Cruz, CA 95064, USA }
\affil{$^2$ Steward Observatory, University of Arizona, 933 North Cherry Avenue, Tucson, AZ, 85721, USA}
\affil{$^3$ Research School of Astronomy \& Astrophysics, Mount Stromlo Observatory, Cotter Road, Weston ACT 2611, Australia}
\affil{$^4$ Korea Astronomy and Space Science Institute, 776 Daedeokdae-ro, Yuseong-gu Daejeon, Korea}
\affil{$^5$ Instituto de Astrofisica, Pontificia Universidad Catolica de Chile, 7820436 Santiago, Chile}
\affil{$^6$ Chinese Academy of Sciences South America Center for Astronomy, 7591245 Santiago, Chile}
\affil{$^7$ National Astronomical Observatory of Japan, Mitaka, Tokyo,181-8588, Japan}
\affil{$^8$ The Kavli Institute for Astronomy and Astrophysics, Peking University, Beijing 100871, P. R. China}
\affil{$^{9}$ Hubble Fellow}

%\affil{$^8$ National Optical Astronomical Observatory, Tucson, AZ, 85719, USA}

\altaffiltext{1} {Email: zcai@ucolick.org}

\begin{abstract}
Cosmological simulations suggest a strong correlation between high optical-depth Ly$\alpha$ absorbers, which arise from the intergalactic medium (IGM), and 3-D mass overdensities on scales of $10-30$ $h^{-1}$ comoving Mpc. By examining the absorption spectra of $\sim$ 80,000 QSO sight-lines over a volume of 0.1 Gpc$^3$ in the Sloan Digital Sky Survey III (SDSS-III), we have identified an extreme overdensity, BOSS1441, which contains a rare group of strong Ly$\alpha$ absorbers at $z=2.32\pm 0.02$. This absorber group is associated with six QSOs at the same redshift on a 30 comoving Mpc scale. Using Mayall/MOSAIC narrowband and broadband imaging, we detect Ly$\alpha$ emitters (LAEs) down to $0.7\times L_{\rm{Ly\alpha}}^*$, and reveal a large-scale structure of Ly$\alpha$ emitters (LAEs) in this field. Our follow-up Large Binocular Telescope (LBT) observations have spectroscopically confirmed 19 galaxies in the density peak. We show that BOSS1441 has an LAE overdensity of $10.8\pm 2.6$ on a 15 comoving Mpc scale which could collapse to a massive cluster with $M\gtrsim10^{15}$ M$_\odot$ at $z\sim0$. This overdensity is among the most massive large-scale structures at $z\sim2$ discovered to date.  %Moreover, we detect a ultraluminous, giant Ly$\alpha$ nebula in the density-peak area. 
\end{abstract}p

\section{Introduction}

Understanding how cosmic fluctuations turn into  structures and galaxies on large scales is one of the fundamental goals 
of observational cosmology. People have observed clusters at $z\sim 2$ with a hot X-ray emitting  intracluster medium (ICM) and a mass of $\sim10^{14}\ M_\odot$ \citep[e.g.][]{gobat11, wang16}, indicating the formation stage of these clusters and their progenitors (e.g., protoclusters) should be earlier than $z=2$ \cite[e.g.][]{blakeslee03, papovich10, fassbender11}. 
 
 The clusters and protoclusters at $z=2$--3 are excellent laboratories for understanding structure formation. Firstly, determining the mass and abundance of the cluster progenitors at $z=2-3$ are crucial for building a complete picture of hierarchical growth in cosmic structures \cite[e.g.,][]{spingel05, chiang13, vogelsberger14, casey16}. Secondly, observations suggest that galaxy properties are closely related to the environment. The fundamental star formation-density and morphology-density relations are in place by $z = 1$ \citep{postman05, smith05} and do not strongly evolve from $z = 1$ to $0$ \citep{delucia12}. Hierarchical galaxy formation models predict that these fundamental relations emerge at earlier epochs (e.g., $z=2-3$, Elbaz et al. 2007).  Thus, overdensities at $z=2-3$ are excellent laboratories to study the emergence of the environmental dependence of galaxy properties (e.g., Peter et al. 2007; Uchimoto et al. 2012). %Inm addition, $z=2-3$ is the most active epoch of the Universe when cosmic star formation rate and QSO activities reach the peak \citep[e.g.,][]{madau98, reddy08}. 
Thirdly, hierarchical structure formation models predict that massive galaxies reside in the overdensities, and a large fraction of their stellar mass was accumulated at $z\sim 2-3$ \cite[e.g.,][]{dickinson03, drory05}. Overdense environments at $z=2-3$ provide us excellent sites to study the assembly and evolution of massive galaxies (Lotz et al. 2011; Uchimoto et al. 2012). In turn, the formation of massive galaxies can be used to strongly constrain the assembly history of clusters \citep{muldrew15}. 

 %the overdense regions at $z=2-3$ are good laboratories to study the complex interactions between galaxies and the intergalactic medium (IGM). Particularly, overdense regions are excellent sites to search for Ly$\alpha$ blobs (e.g., Yang et al. 2009; Prescott et al. 2009) and enormous Ly$\alpha$ nebulae (ELANe) (e.g., Cantalupo14; Hennawi15; Cai et al. 2016b).
 % Utilizing these extended sources, the interactions between the intergalactic/circumgalactic gas and massive galaxies may be revealed through the Ly$\alpha$ emission  \cite[e.g.][]{martin15}.Andreon 2008; Papovich et al. 2010; Rettura et al.2010, 2011; Fassbender et al. 2011b)

In the past few years, people have successfully discovered $\gtrsim20$ galaxy overdensities at $z>2$. There are different approaches to trace the overdense regions at high redshifts, such as with deep redshift surveys of galaxies \cite[e.g.,][]{steidel98,steidel05, chiang14, leeK14}, or with biased halos such as QSOs \cite[e.g.,][]{hu96}; high-$z$ sub-mm galaxies \citep{planck15, casey15, casey16}; and radio galaxies \citep{venemans07}. Nevertheless, current galaxy redshift surveys are often limited by relatively small survey areas (up to a few deg$^2$). Biased halos may have relatively small duty cycles \citep{white12} which make the overdensity selection to be incomplete. Due to these difficulties, people have not constructed a uniform sample at $z>2$ that contains a sufficient number of overdensities/protoclusters for robust comparisons between observations and hierarchical structure formation models \citep{leeK14, chiang13}. This observational challenge motivates us to devise a more complete technique of identifying galaxy overdensities at $z > 2$, especially the most massive overdsensities which place the most stringent constraints on models of structure formation. 

 Cai et al. (2015) developed a novel approach for identifying the extreme tail of the matter density 
distribution on $10-30$ $h^{-1}$ comoving Mpc (cMpc) scales. This approach utilizes the 
largest QSO spectral library from the  Baryon Oscillations Spectroscopic Survey (BOSS) \cite[e.g.,][]{dawson13}, 
%The BOSS project is part of the Sloan Digital Sky 
which enables one to locate extremely rare, coherently strong (high effective optical depth) Ly$\alpha$ absorption due to IGM overdensities on $\sim 15 $ $h^{-1}$ cMpc scale. These IGM HI overdensities, in turn, should trace the most massive early proto-clusters and overdense regions. 
This technique allows the coverage of a larger survey volume and may be less biased, since HI density is correlated with the underlying dark matter density field over large scales (Cai et al. 2015). We named our project MApping the Most Massive Overdensity Through Hydrogen (MAMMOTH) (Cai et al. 2015). 

In this paper, we present the first field, BOSS1441, selected from the early BOSS data release (DR9) using this novel technique. This field contains a group of high optical depth 
Ly$\alpha$ absorption at $z=2.32\pm0.02$, and it is further associated with a QSO overdensity at $z=2.32\pm0.02$. This group of Ly$\alpha$ absorption and QSOs traces an extremely massive overdensity of Ly$\alpha$ emitters (LAEs) at $z=2.32$. In the density peak (on 15 cMpc scale), the LAE overdensity reaches $\approx$ 10.8, making this field one of the most overdense high-$z$ structures discovered to date. %Furthermore, we discover a ultraluminous, giant Ly$\alpha$ nebula in the density peak of BP14 field. This Ly$\alpha$ nebula has a size of $\approx 400$ kpc, larger than typical circum-galactic scale, and shows an extended \civ\ and \heii\ emission, indicating a strong outflow that could be due to AGN-driven wind (e.g., Harrison et al. 2012, Greene et al. 2012, Zakamska et al. 2014). 

 This paper is structured as follows. In \S2, we introduce the SDSS-III/BOSS QSO spectral library and the target selection of the BOSS1441 overdense field. In \S3, we introduce the observations on this field, including the deep narrowband, multiple-wavelength broadband imaging and the multiple-object spectroscopic follow-ups. In \S 4, we describe the observational results. We present the large-scale structure BOSS1441, and we quantify the overdensity using LAEs. Throughout this paper when measuring distances, we refer to comoving separations or distances. We use cMpc to represent comoving Mpc, and pMpc to represent physical Mpc. We convert redshifts to comoving distances assuming 
a $\rm{\Lambda}$CDM cosmology with $\Omega_m= 0.3$, $\Omega_{\Lambda}=0.7$, and $h=0.70$.

\section{Target Selection}

\subsection{SDSS-III/BOSS QSO library}

We used the QSO spectral library observed in the SDSS-III Baryon Oscillation Spectroscopic Survey (BOSS) \citep{dawson13, ahn14} to select fields that contain the group of high optical depth, coherently strong Ly$\alpha$ absorption. 
BOSS is one of the four surveys in SDSS-III, and it is a spectroscopic survey using the 2.5-meter Sloan telescope \citep{gunn06}. 
The BOSS spectra have a resolution of $R\approx 2,000$, with a spectral coverage from 3600 \AA\ $-$ 10,400 \AA\  \citep{bolton12, ross12, dawson13}. For each plate, the BOSS spectra have a typical exposure time of 1-hour, yielding a median signal-to-noise ratio (S/N) of $\sim2$ per pixel at QSOs' rest-frame wavelength $\lambda = 1041- 1185$ \AA\ \citep{lee12}. A pixel in BOSS spectra at $\lambda\approx4000$\AA\  approximately covers 1\AA, which is equivalent to 1 cMpc at $z\approx2.3$. We selected the first field from BOSS DR9 QSO catalog \citep{paris12}. 
The SDSS-III/BOSS DR9 QSO catalog contains $\sim$ 80,000 QSOs over 3,000 deg$^2$, yielding an average  QSO density of 1 per (15 arcmin)$^2$, where (15 arcmin)$^2$ $=$ (17 $h^{-1}$ cMpc)$^2$ at $z \approx 2.3$.  In order to identify absorption systems in the QSO spectra, we use a mean-flux-regulated principal component analysis (MF-PCA) technique to fit the QSO continuum fitting \citep{lee13}. The MF-PCA technique utilizes the PCA fitting to fit the shape of the Ly$\alpha$ forest continuum. The slope and amplitude of the continuum are adjusted using the external constraints of the mean optical depth of the Ly$\alpha$ forest  \citep{lee12, becker13}. 

\subsection{NB403 narrowband filter}

The narrowband filter NB403 centered at $\lambda_c= 4030$\AA, with a FWHM of $\approx 40$ \AA\ (Yang et al. 2008), corresponding to a distance of $\approx 40$ cMpc  along the redshift direction. We constrain our selection within  $z=2.32\pm0.02$ to match our custom narrowband filter {\it NB403}. The {\it NB403} is an ideal filter for this survey: (1) the central wavelength corresponds to Ly$\alpha$ emission $z=2.32$, ideal for seaching overdensities using the BOSS Ly$\alpha$ survey, because the BOSS QSO density peaks at slightly higher redshifts of $z\approx 2.3$. (2) The LAEs at $z=2.3$ is ideal for follow-up ISM/CGM studies: \civ~$\lambda1548/1550$ and \heii~$\lambda1640$ lie in the $V$-band, \ciii$]~\lambda1907/1909$ in the R band, MgII in the $z$-band, and the strong rest-frame optical lines, such as H$\alpha$, H$\beta$, [\oii], [\oiii] in the $Y$ to $K$ bands. 

\subsection{Selection of BOSS1441 field from SDSS-III/DR9}

We select strong (10 -- 30 $h^{-1}$ cMpc) IGM Ly$\alpha$ absorption to trace massive overdensities. 
Following Cai et al. (2015), our new approach utilizes QSO spectra observed from the SDSS-III/BOSS database, 
 allowing us to locate the high optical depth, coherently strong Ly$\alpha$ absorption systems (CoSLAs) on $10 - 30\ h^{-1}$ cMpc scales. These CoSLAs trace the most massive galaxy overdensities at $z\approx2.2-3.5$. At $z<2.4$, the average background QSO density reaches $\ge 20$ per deg$^2$ and this density enables us to select galaxy overdensities using groups of strong Ly$\alpha$ absorption systems.

Cai et al. (2015) proposed that for selection of IGM overdensities $z<2.4$ in the BOSS footprint, the following four criteria can be used: (a) $w_{0.8}< 70$ \rm{\AA}, where $w_{0.8}$ is defined as the width at flux/continuum$=0.8$. (b) The mean flux of the absorption trough ($F_{\rm{trough}})> 0.15$.  (c) Non-detection of low-ionization metal lines associated with the Ly$\alpha$ absorption systems. 
(d)  The presence of a group of absorption systems with $\ge 4$ absorption systems in a volume of ($20\ h^{-1}$ cMpc)$^3$, where each absorption has a $\tau^{15h^{-1} \rm{cMpc}}_{\rm{eff}}$ $\ge 3\times $ mean optical depth $\mean{\tau(z)}$. Cai et al. (2015) suggest that each absorption group contains at least one Coherently Strong Ly$\alpha$ Absorption (CoSLA) which traces the density peak area and has $\tau_{\rm{eff}}^{15h^{-1}{\rm{cMpc}}}\ge 4.5\times\mean{\tau_{\rm{eff}}}$.

We automatically searched for Ly$\alpha$ regions with an optical depth on 15 $h^{-1}$ cMpc scale (\taufif) greater than $3\times \mean{\tau_{\rm{eff}}}$ within $z=2.32\pm0.02$. 
We excluded strong Broad Line Absorptions (BALs) and Damped Ly$\alpha$ Absorption systems (DLAs) using our selection criteria (a) -- (c). We also checked that these absorption systems are not flagged as DLAs in the SDSS-III/BOSS DLA catalog \citep{noterdaeme12}. We then checked the spectra of the nearby QSOs for absorption systems at a similar redshift, and identified fields containing $\ge 4$ absorption systems (in independent QSO sight lines) with \taufif $\ge 3\times\mean{\tau_{\rm{eff}}}$ within 20 $h^{-1}$ cMpc scales. In total, we selected 11 fields that contain such absorption groups. In these 11 fields, we chose our first field with a central coordinate of  $\alpha=14$:41:28.80, $\delta=+40$:01:48 (J2000). This field contains a group of six absorption systems (Figure~1), and the absorption group is further associated with six QSOs residing in the same field and at the similar redshift of $z=2.32\pm0.02$. In Figure~2, we show the Ly$\alpha$ absorption spectra of this group.  We further present the ratio of the \taufif\ to the mean optical depth ($\mean{\tau_{\rm{eff}}}$) at $z=2.3$ (Figure~3). In both Figure~2 and Figure~3, the left two absorption systems (Absorption 1 and Absorption 2) have an effective optical depth higher than the CoSLA theshold ($\tau_{\rm{eff}}^{15h^{-1}{\rm{Mpc}}}\ge 4.5\times\mean{\tau_{\rm{eff}}}$) (Cai et al. 2015). BOSS1441 contains the most CoSLA candidates and QSOs among all the 11 fields. We then selected BOSS1441 field to be the highest priority field in our sample. We adopted a mean optical depth ($\mean{\tau_{\rm{eff}}}$) to be 0.19  at $z=2.3$  \cite[e.g.,][]{becker13}. 
This absorption group, together with QSO group, make this field our highest priority candidate to search for the  large-scale structure. 
We refer to this target as {\it BOSS1441}.

\figurenum{1}
\begin{figure}[tbp]
\epsscale{1.2}
\label{fig:02+04}
\plotone{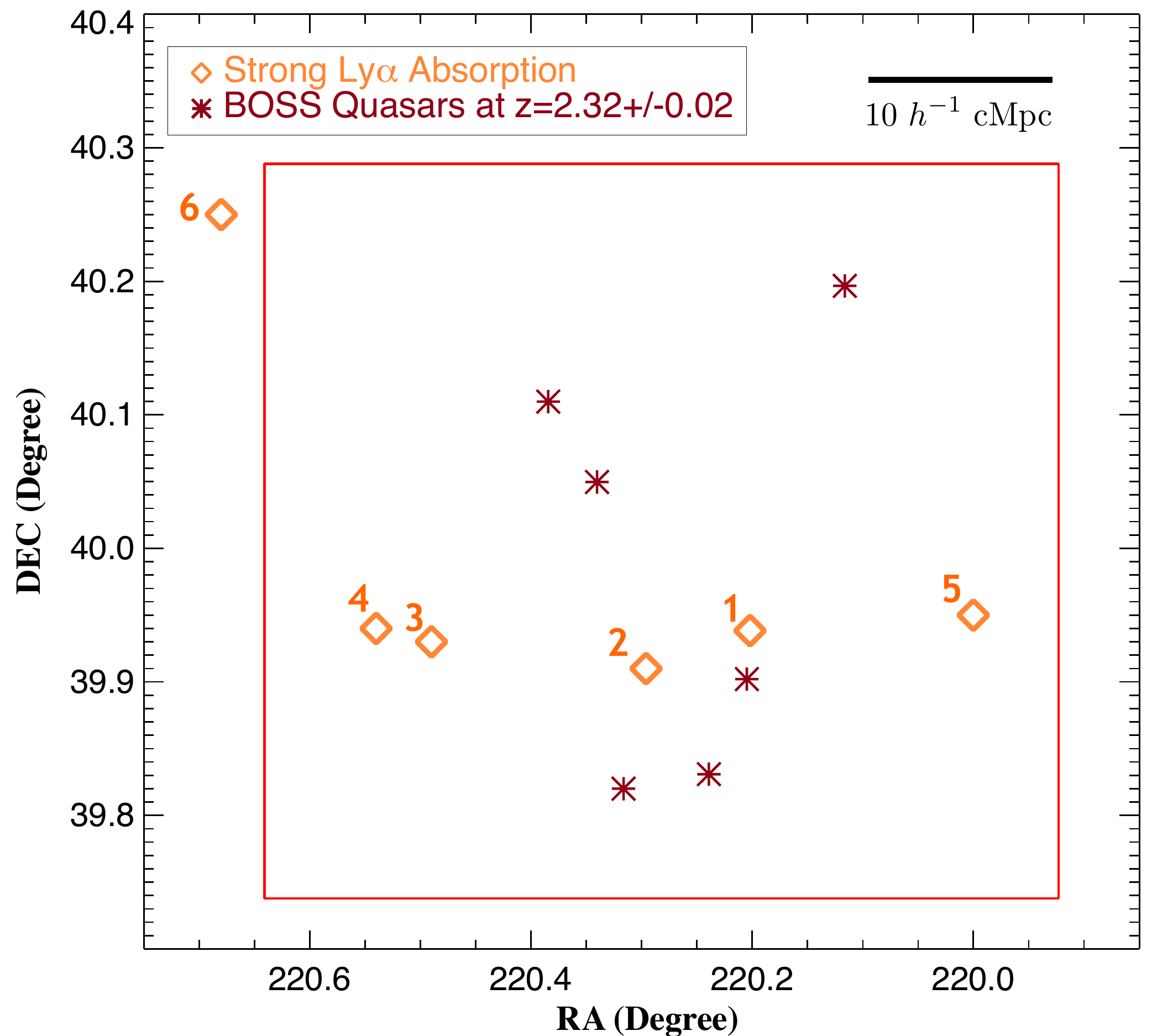}
\caption{BOSS1441 field selected from the 3,000 deg$^2$ SDSS-III/BOSS DR9 survey. This field contains a group of strong Ly$\alpha$ absorption systems (orange diamonds) and QSOs (brown asterisks) over 30 $h^{-1}$ cMpc at $z=2.3$.  Each Ly$\alpha$ absorption is presented in Figure~2 and Figure~3. }
\end{figure}

\figurenum{2}
\begin{figure}[tbp]
\epsscale{1.2}
\label{fig:02+04}
\plotone{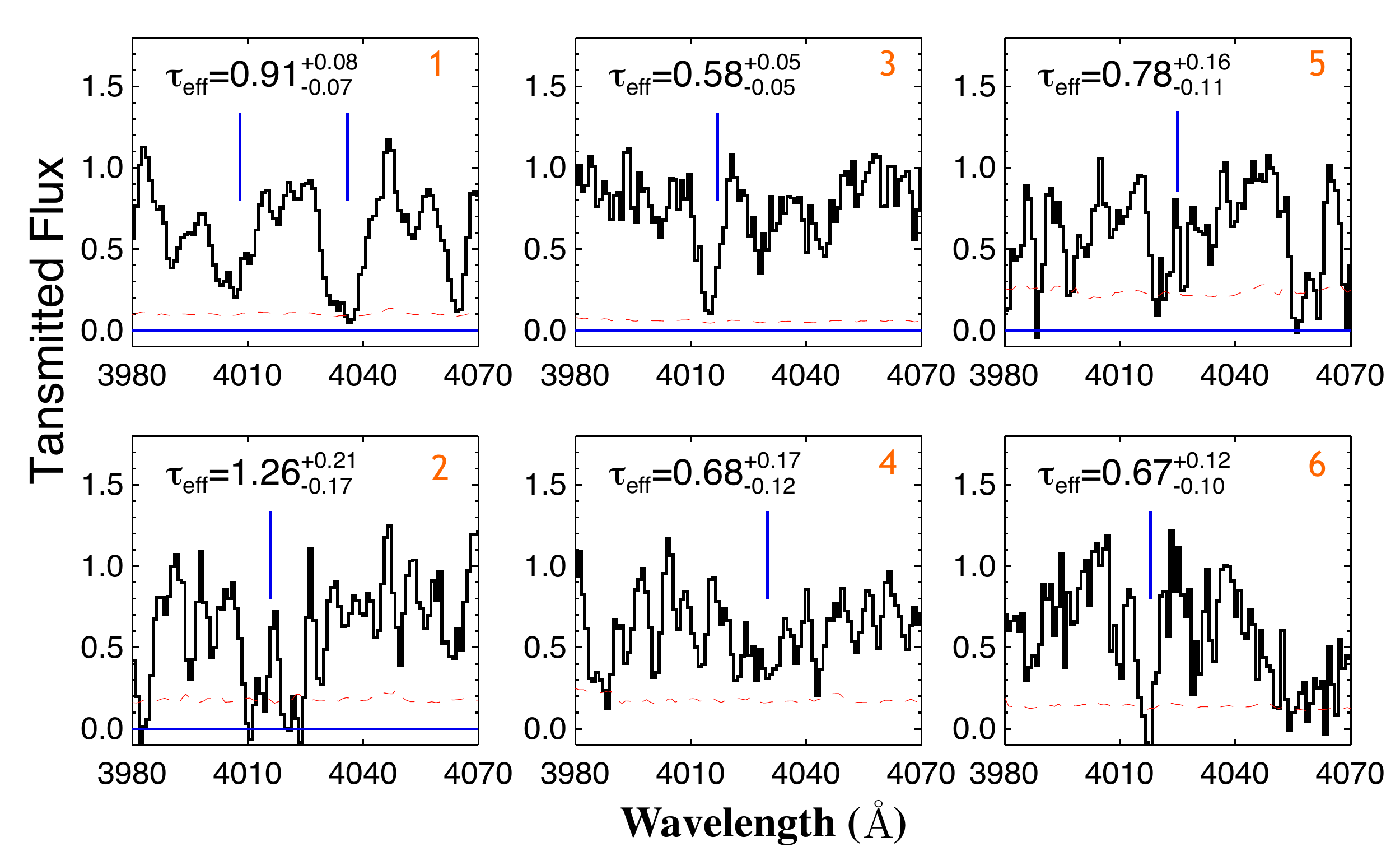}
\caption{Snippets of the absorption group in the field of BOS1441. We present the continuum-normalised QSO spectra at Ly$\alpha$ absorption at $z=2.3$ in each panel. In each spectrum, we detect deep Ly$\alpha$ 
absorption (blue bar) at the same redshift as the QSO group. The errors are presented using red dashed line. The mean optical depth at $z=2.3$ is $\approx$0.19 \cite[e.g.,][]{becker13}. The left two panels are two Coherently Strong Ly$\alpha$ Absorption Systems (CoSLAs) candidates with the $\tau_{\rm{eff}}$ over 15 $h^{-1}$ cMpc scale greater than 4.5$\times \mean{\tau_{\rm{eff}}}$ (see Figure~3).  }
\end{figure}

\figurenum{3}
\begin{figure}[tbp]
\epsscale{1.2}
\label{fig:02+04}
\plotone{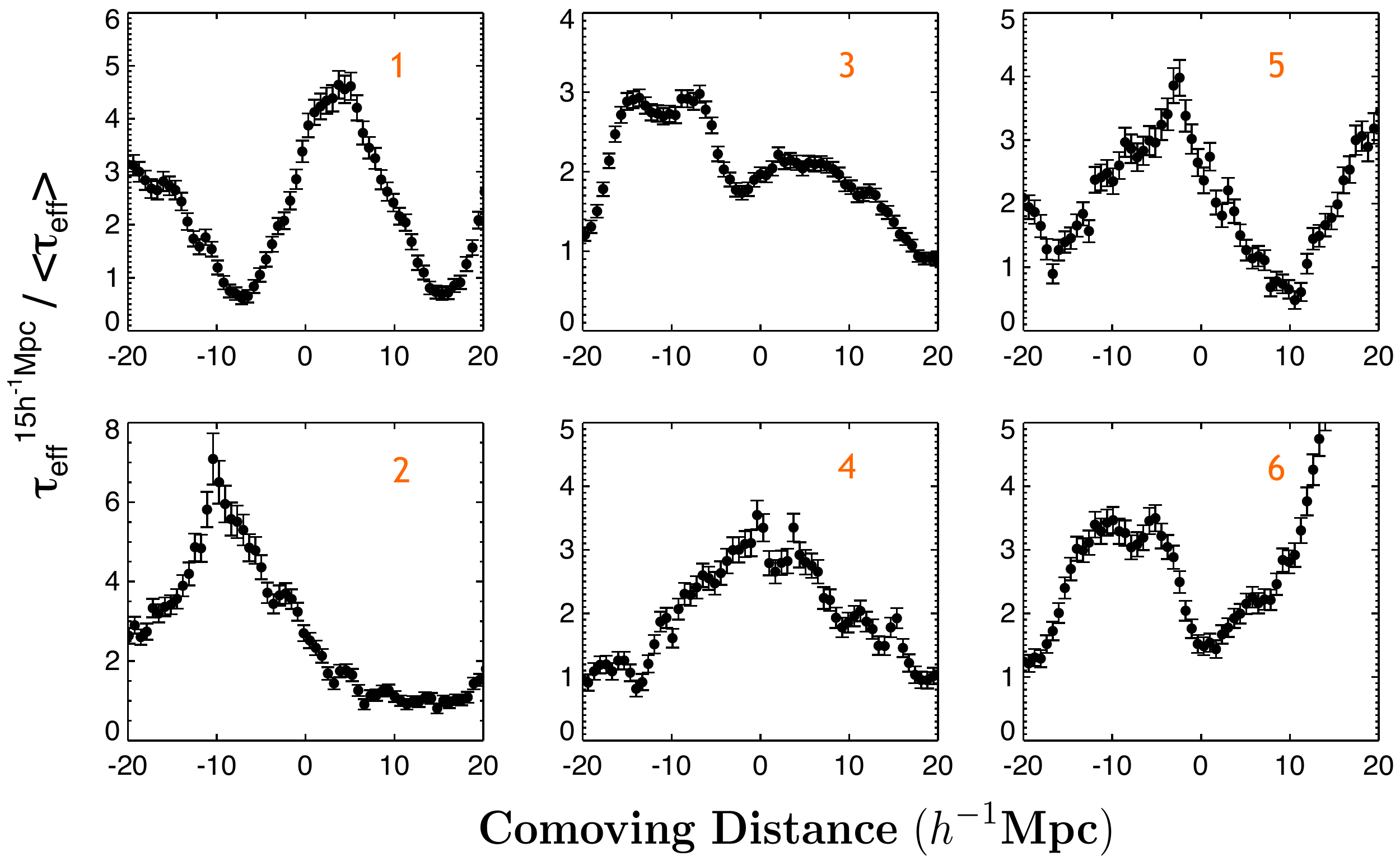}
\caption{Effective optical depth over 15 $h^{-1}$ cMpc ($\tau^{15h^{-1}\rm{Mpc}}_{\rm{eff}}$) in the BOSS1441 field centered at $z=2.32$ for each QSO spectrum in Figure~2. In each spectrum, we see strong Ly$\alpha$  absorption with high optical depth. This group of Ly$\alpha$ absorption traces the large-scale structure BOSS1441.  }
\end{figure}

\subsection{Survey Volume}

BOSS1441 is selected from SDSS-III/DR9. In the SDSS-III/DR9 database, we chose fields that can be observed in the spring semester from RA = 9-hour to RA = 16-hour.  We selected fields that contain $\ge 4$ background  QSOs ($z_{\rm{em}}>2.34$) within a $20\ h^{-1}$ cMpc $\times$ $20\ h^{-1}$ cMpc area. We have 2622 independent fields that contain groups of background QSOs.  The survey volume of the 2622 groups is $2622\times 20\ h^{-1}$ cMpc $\times 20\ h^{-1}$ cMpc $\times 44$ cMpc $ \approx 1.0\times10^8$ cMpc$^3$. Thus, BOSS1441 can be regarded as the most extreme system selected from a  $ \approx 1.0\times10^8$ cMpc$^3$ survey volume using our selection technique. 

\section{Deep imaging and spectroscopy in BOSS1441 field}

After selecting the BOSS1441 field, we conducted deep narrowband, broadband imaging and the multiple-slit spectroscopy to search for Ly$\alpha$ emitters (LAEs) in this field. %We also use broad-band selected galaxies at $z\approx2.3$ (BX-galaxies) \citep{adelberger05} to map out the overdensity and quantify its strength. 

\subsection{KPNO-4m/MOSAIC narrowband + broadband imaging}

We used the MOSAIC1.1 camera \citep{sawyer10} 
at the f/3.1 prime focus of the 4m Mayall Telescope
at the Kitt Peak National Observatory to conduct deep narrowband and broadband imaging. 
 We obtained deep narrowband images with the {\it NB403} narrowband filter. As shown in \S 2.2, NB403 is used for probing the Ly$\alpha$-emitting galaxies at $z\approx 2.32$. 
The BOSS1441 overdensity was observed on Mar 2013, Apr. 2014, and Jun. 2014. The total exposure time is 17.9 hours, which consists of individual 15 or 20 
minute exposures with a standard dither pattern to fill in the
gaps between the eight chips. The seeing ranges from $1.1''- 1.7''$, with a median combined seeing of 1.32$''$. We observed with the broadband filter (Bw) for 3-hr to match our selection criteria.  These observing conditions enable us to achieve a narrowband magnitude $m_{\rm{NB403}}=25.1$ at 5-$\sigma$ using a circle aperture of 2.5$''$ and a Bw-band magnitude $m_B=25.9$ at 5-$\sigma$ using the same aperture. The narrowband sensitivity limit corresponds to a $1\sigma$ Ly$\alpha$ surface brightness of  SB$_{\rm{Ly\alpha}}= 2.4\times10^{-18}$ erg s$^{-1}$ cm$^{-2}$ arcsec$^{-2}$. 

\subsection{LBT multiple-wavelength broadband imaging}

Our multi-band imaging is conducted by the Large Binocular Cameras (LBC) \citep{giallongo08} at the LBT on Mount Graham in Arizona. LBC is characterized by a unique optical system. Each of the two mirrors has a collecting area of $8.4\times8.4$m$^2$ and a field of view of 23' $\times$ 23'. 

For a complete selection of star-forming galaxies at $z\approx 2.3$, 
we used the LBT to conduct deep multiple broadband observations in the $U$, $V$, and $i$ filters \citep{steidel05}. The multiple broadband imaging can also eliminate the low-redshift [\oii] interlopers which are the main contaminants  to LAEs at $z=2.3$. We use the standard $z=2$ star-forming galaxy selection (BX selection) technique \citep{adelberger05} to eliminate the low-$z$ interlopers. 
The one-night LBT/LBC deep 
multi-color imaging was carried out in Mar. 2014. The LBC binocular mode was used, and the LBC blue used solely for deep $U$-band observations. On the red side, we observed the same field in the $V$-, $i$-filters. We conducted 6-hour deep imaging in $U$, 2.5-hour in $V$, and 3.5-hour in $i$. The individual exposure was six minutes with a standard 5-point dither pattern to fill in the CCD gaps.  The conditions were partially cloudy, with a median seeing of 0.6$''$. Our observations reached $U_{\rm{AB}}= 26.5$ at 5-$\sigma$, $V_{\rm{AB}}=26.4$ at 5-$\sigma$ and $i_{\rm{AB}}=26.4$ at 5-$\sigma$. These depths are deep enough for selecting $0.5\times L^*$ galaxies (e.g., Adelberger et al. 2003).

\subsection{LBT multiple-slit spectroscopy}

We used the Dual channel of  Medium-Dispersion Grating Spectroscograph (MODS) \citep{byard00} to spectroscopically confirm of galaxies in the overdense field. The LBT/MODS has a field of view of 6' $\times$ 6' for each blue and red channel, and it provides high efficiency in the full wavelength range of 3200\AA\ -- 10,000\ \AA\ with $R= 2000$ for most of the slits. 
We used a dichroic that divides the incoming beam at $\approx$ 5700\AA. This configuration covered Ly$\alpha$ and the interstellar lines, such as \civ, \heii, \ciii], for LAEs at $z=2.3$.

On May 30 and May 31 2014, we conducted deep spectroscopy using two masks of LBT/MODS. Each masks containing 15 candidates of LAEs and BX galaxies. 
The first mask was anchored by an enormous Ly$\alpha$ nebula (ELAN) (see \S4.1). We observed this mask for a total of six hours. The second mask was observed over four hours. For each mask, we divided the total exposure into multiple individual 1,800 sec integrations. The average seeing for the MODS observations is 1$''$. 
The MODS data reduction follows the LBT/MODS reduction routine (Croxall et al. 2015).   Each raw image was processed with the MODS CCD reduction utilities (modsTools v0.3) to obtain bias-subtracted and
 flat-fielded images. We fit polynomial the arc calibration with polynomials to determine the transformations between image
pixels and wavelength. The sky model was fit to each image using B-splines and then subtracted. 
The LACOSMIC routine (Van Dokkum 2010) was used to identify the cosmic rays and to reject them during the construction of the sky model. The error in 
the wavelength calibration due to the telescope flexure was corrected using sky lines. %The
%individual exposures were combined with inverse variance
%weighting to produce the final 2D spectrum.

\figurenum{4}
\begin{figure}[tbp]
\epsscale{1.2}
\label{fig:02+04}
\plotone{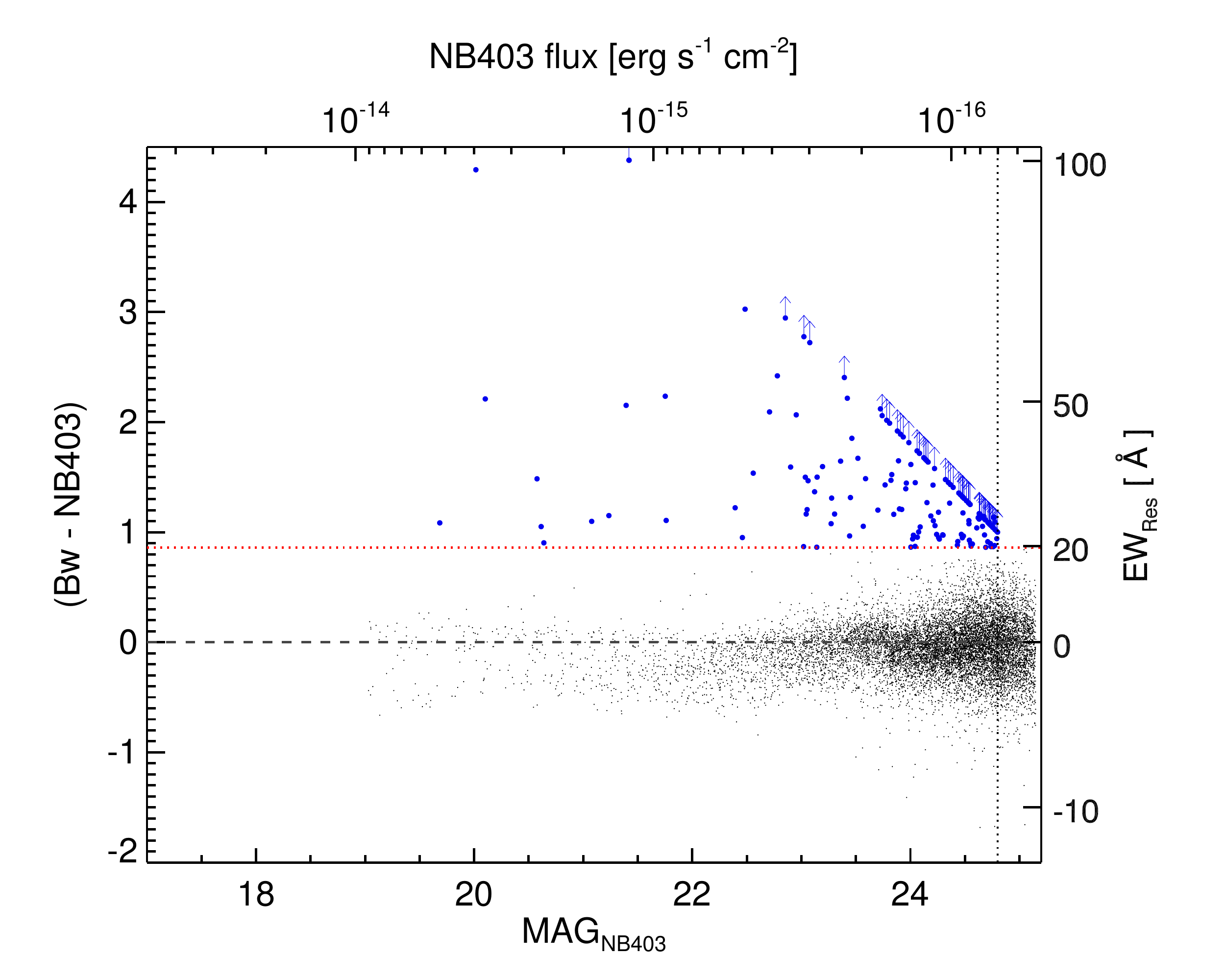}
\caption{The ($Bw-NB403$) color-magnitude diagram for all sources detected in  BOSS1441. The narrowband data reach $NB403=$24.8 at the 5$\sigma$ level. We select line-emission objects (blue points) with the criteria NB $<$ 24.3 (vertical black line) and $Bw-NB403>0.87$ (red dashed line; equivalent to rest-frame EW $>$ 20\AA). } %We detect confirm a LAE overdensities at the bright end: 82 LAEs over a large scale of 35 $h^{-1}$ Mpc satisfy the color selection criterion, whereas 42 LAEs are expected in a random field based on luminosity function calculations (Guaita et al. 2010, Ciardullo et al. 2012). }
\end{figure}

\figurenum{5}
\begin{figure}[tbp]
\epsscale{1.2}
\label{fig:02+04}
\plotone{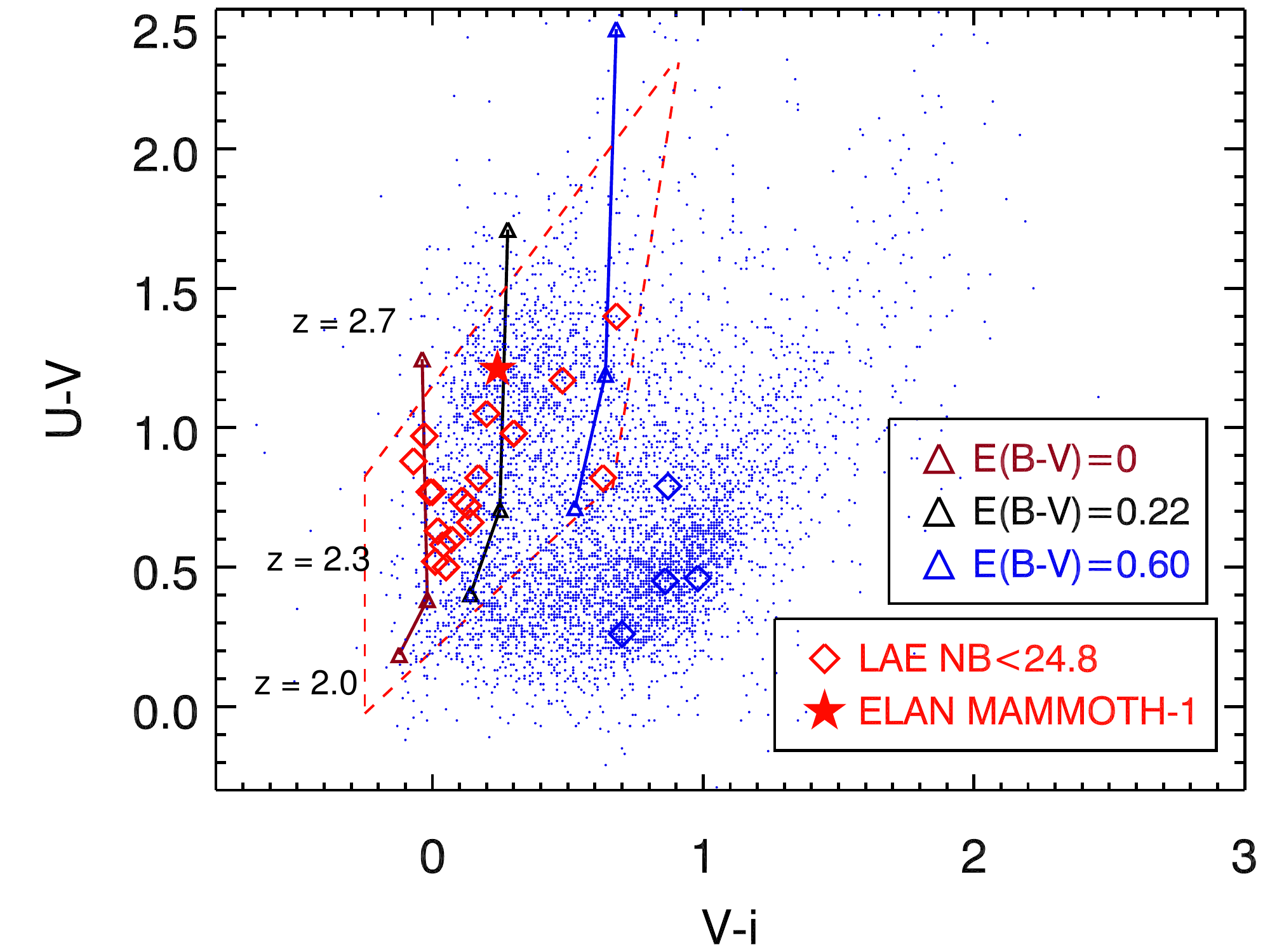}
\caption{The broadband color of the Ly$\alpha$ emitters (LAEs) in the density peak region of 15 cMpc$^3$. The broadband colors in $U$, $V$ and $i$ are obtained using Large Binocular Camera (LBC) on the Large Binocular Telescope (LBT). The dotted red lines outline the selection region. Most of the LAE candidates have broadband color consistent with the $z=2.0$-- 2.7 galaxies (BX galaxies). There are four LAE candidates (blue diamonds) locate at the low-$z$ galaxy locus and we exclude these four galaxies from our LAE sample. }
\end{figure}

\figurenum{6}
\begin{figure*}[tbp]
\epsscale{0.95}
\label{fig:overdensity}
\plotone{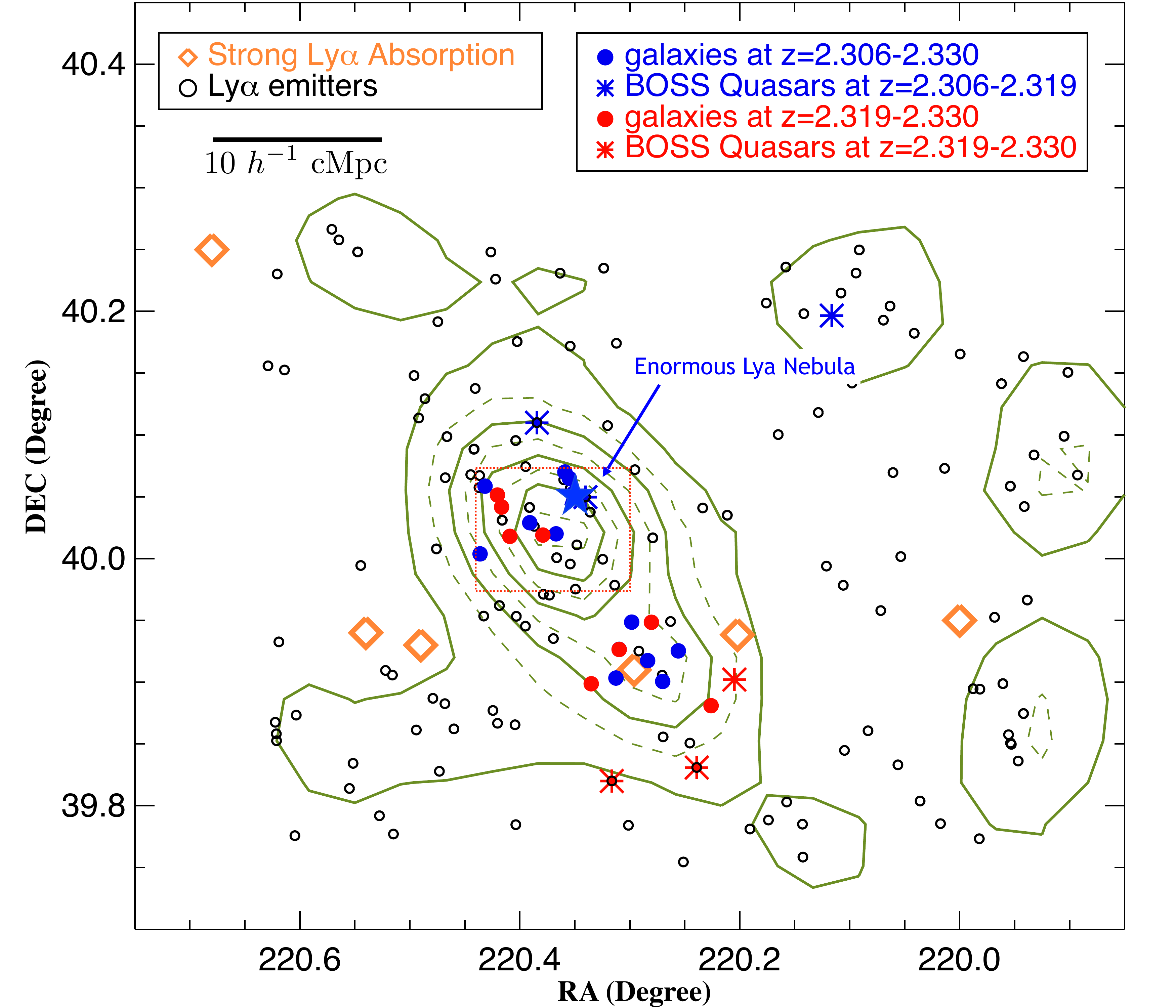}
\caption{The galaxy overdensity BOSS1441 at $z=2.32\pm0.02$, selected from the SDSS-III/BOSS DR9 database. This structure is traced by a group of Ly$\alpha$ absorption systems (orange diamonds) and QSOs (brown asterisks) over 30 $h^{-1}$ cMpc at $z=2.3$.  Each Coherently Strong Ly$\alpha$ Absorption (CoSLA) candidate is presented in Figure~2 and Figure~3. %In the whole MOSAIC field (72,000 Mpc$^3$ volume), the LAE density is a factor of $2.0\times$ the random field (Ciardullo et al. 2012). 
Our LBT/MODS spectroscopy have targeted 20 LAEs with mag$_{\rm{NB403, AB}}<24.8$ with two masks in 
an area of 50 arcmin$^2$ (8 $h^{-1}\times 8^{-1}$) cMpc$^{2}$, and confirmed 19 of them (blue and red circles). The LAE density in the density peak region of (15 cMpc)$^3$ volume (red dotted box) is $11.8\times$ that in random fields ($\delta_g=10.8$). The green density contours are overplotted to show the density map of LAEs. Each contour (solid and dashed) indicates an increase in the galaxy number density of 0.1 galaxies per arcmin$^2$. We determine the redshift of each spectroscopically-confirmed LAE based on the Ly$\alpha$ emission. }
\end{figure*}

\section{Results}

 %We have developed algorithms to select high optical depth systems that are due to IGM overdensities rather than discrete damped Ly$\alpha$ absorption systems (DLAs). %Left panel of Figure 1 demonstrates that, from the simulation, the overdensities traced by largest IGM Ly$\alpha$ absorption are generally more massive than that traced by other rare sources (e.g., quasars). Right panel of Figure 1 gives an example of the largest 1-D IGM Ly$\alpha$ absorption selected from early data release of SDSS-III, which could trace an underlying massive overdensity. 
%\end{abstract}

{\subsection{Discovery of a large-scale structure BOSS1441}}

Our narrowband imaging has a field of view of $35'\times35'$ ($41\ h^{-1}\times 41\ h^{-1}$ cMpc$^2$ at $z=2.3$), and a redshift range of $2.32\pm0.02$. 
In Figure~4, we present the color selection criteria for the LAEs. We use the color criterion of {\it Bw} $- NB  >0.87$, corresponding to a Ly$\alpha$ rest-frame equivalent width of EW$_{\rm{Ly\alpha}}$ $>20$ \AA. We cut our selection of LAEs down to a narrowband magnitude of $m_{\rm{NB403,AB}}=24.8$, where the LAE selection completeness is $\ge 90\%$ \cite[e.g.,][]{guaita11, zheng16}. This limit corresponds to a Ly$\alpha$ luminosity of $1.56\times10^{42}$ erg s$^{-1}$,  $0.73\ L^*_{\rm{Ly\alpha}}$, where $L^*_{\rm{Ly\alpha}}= 2.1\times10^{42}$ erg s$^{-1}$ \citep{ciardullo12}. The total volume of the MOSAIC imaging is $41\times 41 \times 32 \ h^{-3}$ cMpc$^3$. Over this volume,  we detect 99 LAE candidates that satisfy the color criterion of  {\it Bw} $- NB  >0.87$. 

We use LBT/LBC to check the broadband colors of the LAE candidates. We use the $U-V$ vs. $V-i$ colors of the LAE candidates.  Following Adelberger et al. (2003), our selection criteria (red trapezoid in Figure 5) were found by running Bruzual \& Charlot (2003, hereafter BC03) models. We use a galaxy template with 0.2 $\times Z_\odot$, a constant star formation for 100 Myr, reddened by applying a Calzetti attenuation curve (Calzetti et al. 2000) with E(B $-$ V) = (0, 0.22, 0.6). %We also add the photometric errors of 0.2 magnitude.  

We find that BOSS1441 has a density peak at $\alpha=14$:41:26.40,  $\delta=+40$:01:12.0. For a comparison to previous protoclusters (see Table~2), we count LAEs in a (15 cMpc)$^3$  volume centered on this peak.  In that 6 arcmin $\times$ 6 arcmin area (red dotted box in Figure~6) \footnote{We choose 6 arcmin $\times$ 6 arcmin, because we want to compare the volume in a 15 cMpc $\times$ 15 cMpc $\times$ 15 cMpc box. A 6-arcmin corresponds to $\approx$ 9.5 cMpc. The volume is  9.5 cMpc $\times$ 9.5 cMpc $\times$ 40 cMpc (comoving distance along the sight-line direction)=  (15 cMpc)$^3$, which has the same volume to the overdensity calculations in other fields \cite[e.g.][]{chiang14}.}, we detect 21 objects that satisfy the color criteria $Bw -NB>0.87$; and 15 have broadband colors in $U-V$, $V-i$ consistent with galaxies at $z\sim2.3$. We present the broadband colors of these LAE candidates in Figure~5 (red diamonds). One of the 21 sources is a BOSS QSO at $z=2.31$ (QSO3 in Table~4). 
Four out of the 21 candidates are likely to be low-$z$ [\oii] emitters. Another one out of the 21 candidates have $Bw> 25.9$ and $i>26.5$. Therefore, we have 15 LAE candidates and one BOSS QSO in the density peak region (red dotted box in Figure~6). Among these LAEs, we  detect an enormous Ly$\alpha$ nebula (ELAN). We named this ELAN {\it MAMMOTH-1} (blue star in Figure~6). The size of MAMMOTH-1 is extremely large, about $450$ physical kpc (pkpc) and among the largest ELANs discovered to date. MAMMOTH-1 also contains the strongly extended He II $\lambda1640$ and CIV $\lambda1549$ emission. We will discuss the MAMMOTH-1 nebula in our following paper (Cai et al. 2016b). 

Our color criteria have a high successful rate for selecting galaxies at $z\approx 2.3$. In our LBT/MODS footprint ($50$ arcsec$^2$, covering the density-peak area), we put 20 sources in MODS masks that satisfy both the LAE and BX color selection criteria. We spectroscopically confirm the Ly$\alpha$ emission for 19 of them with $m_{\rm{NB}}<24.8$. 11 of the 19 LAEs reside in the density peak area (red box in Figure~6). Including the BOSS QSO in this region, we have Ly$\alpha$-based redshifts for 20 galaxies in our MODS footprint and 12 galaxies in the density peak region. 
 We summarize the spectroscopically confirmed galaxies in Table~3, and present their Ly$\alpha$ emission lines in Figure~7. We did not detect Ly$\alpha$ emission for only one galaxy candidate, because the expected Ly$\alpha$ line falls in the gap between two CCDs of LBT/MODS.  The spectroscopic results demonstrate that our LAE selection is solid.  

%In random fields, we expected to detect $\sim 1.53$ LAEs in the same volume (Table 1). 
The main low-$z$ interlopers are strong [\oii] emitters at $z = 0.081\pm0.005$ with EW$_{\rm{[OII]}}>60$ \AA. %From Hogg et al. (1998), the number density of such \oii\ interlopers should be low. 
We excluded them before conducting spectroscopic follow-ups, because their broadband colors do not satisfy high-$z$ LAE criteria.  Our spectroscopic survey has a resolution of $R=2000$ and is able to resolve the [\oii]$\lambda\lambda$3727,3729 doublets. Our optical range covers [\oiii], H$\beta$, and H$\alpha$ lines of the emitters. Throughout the spectroscopic check, we did not find any \oii\ interlopers that contaminate our LAE spectroscopic sample.

\subsection{LAE overdensity in the density peak of BOSS1441}

Stark et al. (2014) find that the characteristic size of protoclusters (i.e., defined as progenitors of massive $z=0$ clusters) is about 15 cMpc ($\sim10\ h^{-1}$ cMpc). \citet{steidel98,  steidel05} and \citet{chiang13} estimate the relation between the protocluster mass $m$ and galaxy overdensity $\delta_g$ in a volume of (15 cMpc)$^3$, where $\delta_g$ is defined as $\delta_g= \frac{N_{\rm{group}}}{N_{\rm{field}}}-1$, and $N_{\rm{field}}$ is the mean number of LAEs in random fields. $N_{\rm{group}}$ is the LAE number within the overdensity. For a comparison to previous confirmed overdensities (Table~2), we measure the galaxy overdensity in the density peak of an asymmetric box with a total volume of $15^3$ cMpc$^3$ (red box in Figure~6).

 In Figure~8, we present the radial density profile (black dots with error bars) measured in circular annuli centered on the field center of BOSS1441 ($\alpha=14$:41:26.40,  $\delta=+40$:01:12.0). The $x$-axis is the diameter of the circle. The average overdensity of this protocluster is $>10$ over a large scale of $\gtrsim15$ cMpc. 
 From our LBT/MODS spectroscopy, the LAEs in this region have $z=2.319\pm0.013$. 
According to the previous section, in the density peak of a (15 cMpc)$^3$ volume (red box, Figure~6), we detected 16 LAEs (including a BOSS QSO) down to $L_{\rm{Ly\alpha}}\ge 0.73\times L^*_{\rm{Ly\alpha}}$. 12 out of these 16 LAE candidates are spectroscopically confirmed (including a BOSS QSO). According to previous section, our LAE selection is solid. If the other LAE candidates that we did not put in LBT/MODS masks are real, and considering a 90\% of LAE completeness \citep{guaita11, zheng16}, the intrinsic LAE number should be 17.8 in the density peak. 
From the luminosity function of LAEs at $z=2.1$ \citep{ciardullo12}, the average LAE number is $N_{\rm{field}}=1.53$ in a cube with 15 cMpc on a side. Then, we calculate that the LAE overdensity has a  $\delta_{g}\approx 10.8$, making this field one of the most overdense structure of LAEs ever discovered (see Table 2).

Let us measure the uncertainties of galaxy number counts between different (15 cMpc)$^3$ cells. 
Similar with \citet{leeK14}, we account for both Poissonian shot
noise and sample variance. 
The total fractional errors of $\frac{N_{\rm{group}}}{N_{\rm{field}}}$ is a quadratic sum of
these two error sources:  (1) Shot noise $\delta N_{\rm{shot}}$. Using the Ly$\alpha$ luminosity function \citep{ciardullo12}, in random fields, $N_{\rm{field}}$ is 1.53 within a (15 cMpc)$^3$ box. Thus, the fractional error from shot noise is $\delta N_{\rm{shot}}=\sqrt{17.8}=4.2$, and $\frac{\delta N_{\rm{shot}}}{N_{\rm{field}}}=4.2/1.53= 2.7$. (2) Sample variance ($\delta N_{\rm{SV}}$). $\delta N_{\rm{SV}}$ is determined 
by putting random boxes in different positions with (15 cMpc)$^3$, and measure the variance of the halo counts. 
 The sample variance is measured to be $\frac{\delta N_{\rm{SV}}}{N_{\rm{field}}}=$ 52\% \citep{leeK14, chiang13}. Therefore, we measure the total fraction error  of $\frac{\delta N_{\rm{group}}}{N_{\rm{field}}} = \sqrt{
(\frac{\delta N_{\rm{shot}}}{N_{\rm{field}}})^2+ (\frac{\delta N_{\rm{SV}}}{N_{\rm{field}}})^2}$= $\sqrt{2.7^2+0.53^2}\approx2.7$, which means that the root-mean-square (rms) fluctuation of LAE number per (15 cMpc)$^3$ cell is 
roughly $1.5\times2.7=4.0$, i.e., the measured error of LAE number per (15 cMpc)$^3$ cell is 4.0. 
Therefore, the galaxy overdensity ($\delta_g$) in the density peak is $\delta_g= 10.8\pm2.6$.

\figurenum{7}
\begin{figure*}[tbp]
\epsscale{1.2}
\label{fig:02+04}
\plotone{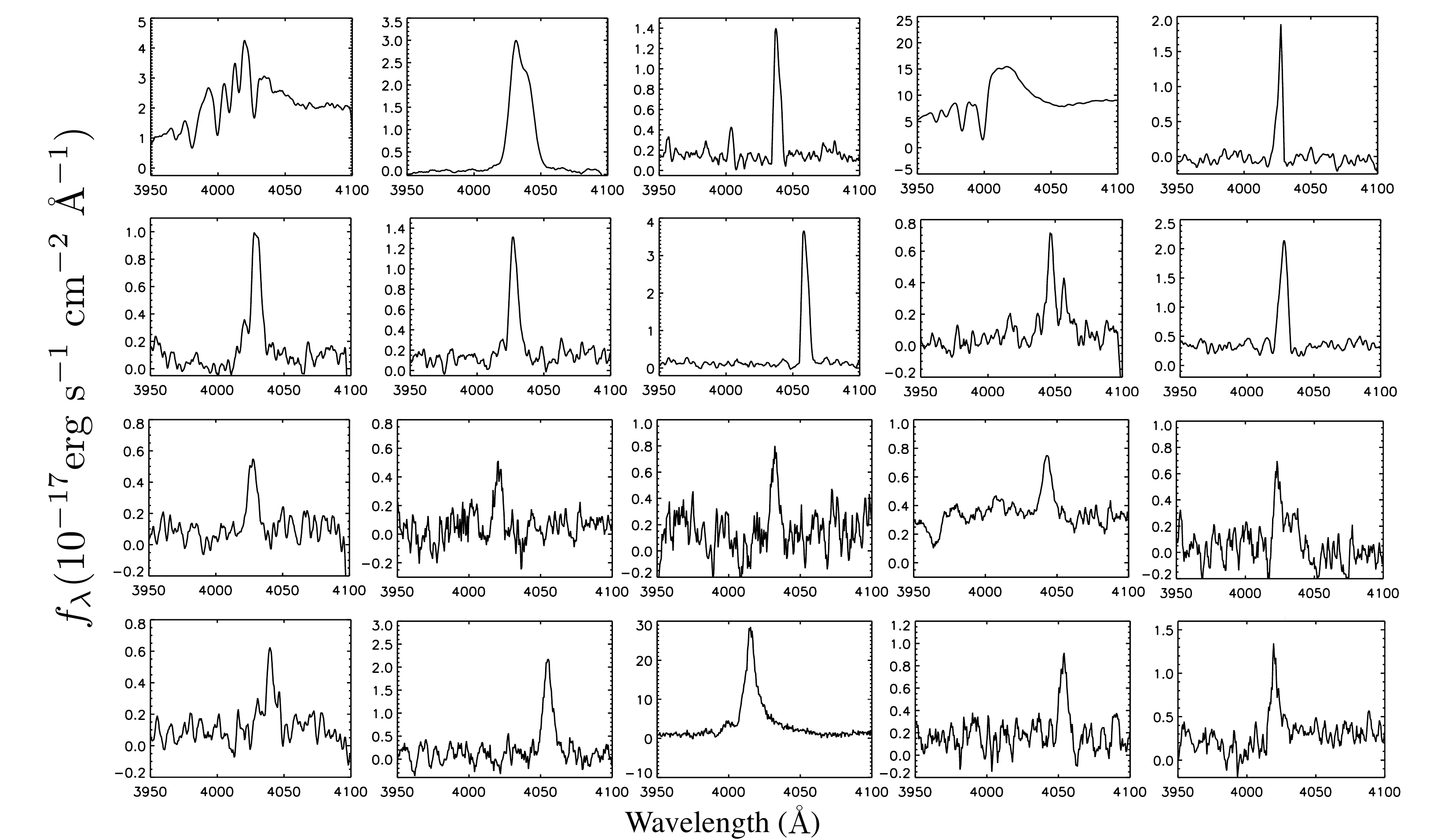}
\caption{Extracted 1-dimensional LBT/MODS spectra for member galaxies of the BOSS1441 overdensity, including one BOSS QSO. All the sources have Ly$\alpha$ line detections with SNR $>10$. }
\end{figure*}

\section{Discussion}

\subsection{Present-day mass of BOSS1441}

If BOSS1441 collapses to a cluster at $z\sim0$, then the collapsed cluster mass can be estimated from the overdensity of galaxies at $z=2-3$ \cite[e.g.,][]{chiang13}.  We measure the mass based on the estimation by \citet{steidel98, steidel05}  and \citet{chiang13}. The progenitors of massive $z=0$ clusters have a characteristic size of 15 cMpc (e.g., Stark et al. 2014). As shown in the previous section, BOSS1441 has an overdensity of $\delta_g=10.8$ over a 15 cMpc scale. If the matter on a 15 cMpc scale collapses to a bound, virialized system at $z\sim0$, the virialized mass ($M_{\rm{z\sim0}}$) can be expressed as:  

\begin{equation}
\rm{M}_{z=0}= (1+ \delta_m) \mean{\rho} V_{\rm{true}}= 2.6\times10^{14}\ (1+\delta_m)/C\ h^{-1}\ \rm{M}_\odot,
\end{equation}
 where  $V_{\rm{true}}$ is the true volume after correcting for redshift-space distortion. $V_{\rm{true}}= V_{\rm{app}}/C$, and $V_{\rm{app}}= 15^3 $ cMpc$^3$, the observed comoving volume in which the measurement is made (within the $6'\times 6'$ region on the plane of sky and neglecting the peculiar velocities between $z=2.31-2.33$). $C$ is a volume factor for correcting the redshift-space distortion caused by peculiar velocities and defined as $\frac{V_{\rm{app}}}{V_{\rm{true}}}$. $\mean{\rho}$ is the mean matter density of the universe, which is equal to $\frac{3H_0^2}{8\pi G}\Omega_m= 8.4\times10^{10}\ h^2\ M_\odot$ cMpc$^{-3}$. The $\delta_m$ is the matter overdensity. The relation between the observed galaxy overdensity and matter overdensity can be expressed as $1+b\delta_m= C(1+\delta_{\rm{g}})$, where $\delta_{\rm{gal, obs}}$ is the observed galaxy overdensity  we measured in \S4.2 and the galaxy bias $b$  could range from $1-4$ (e.g., Steidel et al. 1998). 
Dey et al. (2016) found that LAE is a biased tracer of the matter distributions in an overdensity. They observe that Ly$\alpha$ emission is clearly enhanced in denser regions in their protocluster. We take the LAE bias as $b_{\rm{LAE}}=2.1$ \cite[e.g.,][]{gawiser07, leeK14}. The volume correction factor $C$ can be approximated by an expression of $C= 1+f-f(1+\delta_m)^{1/3}$ \citep{steidel98, steidel05}, where $f= \Omega_m (z)^{0.6}$. For $\delta_g=10.8$ at $z=2.32$, we find $C= 0.56$ and $\delta_m= 2.2$. 
According to the theory of density-perturbation, the overdensity $\delta_m=2.2$ corresponds to a linear overdensity $\delta_L \sim 0.83$ at $z=2.3$ (Eq.18 of \citeauthor*{mo96}). Following the growth of linear perturbation ($\delta_L\propto t^{2/3}$), $\delta_L$ should pass the threshold of $\delta_L> 1.69$ at $z<0.5$. $\delta_L> 1.69$ corresponds to the density contrast of the gravitational collapse of a spherical perturbation \cite[e.g.,][]{jenkins01, reed03}. Thus, we expect that BOSS1441 should collapse to form a cluster at low redshift. Using cosmological simulations, \citet{chiang13} find that $\ge80\%$ of the overdensities with $\delta_g>2.88$ at $z\approx 2$ are protoclusters which will collapse to virialized clusters at $z\sim0$. 

Adopting an LAE bias of $b=2.1$ and a correction factor $C=0.56$, we estimate that BOSS1441 at $z=2.3$ should evolve into a $z=0$ cluster with a mass of $M_{\rm{z\sim0}}\approx 1.5\times10^{15}$ M$_\odot$. Such a high mass  makes this overdensity one of the most massive cluster progenitors (Table 2). In addition to the analytical estimations, \citet{chiang13} calculated the $\delta_g$-$\delta_m$ relation. From Chiang et al. (2013), we estimate that the most massive $M_{\rm{z=0}}\ge10^{15}\ M_\odot$ cluster has a progenitor at $z\approx2.5$ with $\delta_g\approx 5-10$ and $\delta_m\approx2-4$, assuming a galaxy bias $b=2$. In Figure~8, we further compare the radial density profile of BOSS1441 (black points)  with the predicted radial density profiles of cluster progenitors  (Chiang et al. 2013), using a galaxy bias of $b=2$. We find that the radial profile of BOSS1441 matches that of Coma cluster progenitors on scales $\sim 5 - 20$ cMpc. On smaller scales of $<5$ cMpc, the observed galaxy number density is significantly lower than that of Coma cluster progenitors in cosmological simulations. This discrepancy may  indicate that the galaxies in the density peak are not be completely traced by the LAE population \cite[e.g.][]{muldrew15}. In future, we need to conduct a complete galaxy survey from optical to infrared, and study whether the LAE is a good tracer in the density peak, and whether the quenched galaxies exist in the density peak (Feldmann \& Mayer 2014).  

In Cai et al. (2015), we showed the $\delta_m$-$\tau$ relation from the cosmological simulations. 
BOSS1441 overdensity is our first observed field that is traced by a Ly$\alpha$ absorbers. It is interesting to check if BOSS1441 follows the $\delta_m$-$\tau$ relation in the simulation. In the LyMAS simulation (Cai et al. 2015), we found that the mass overdensities can be traced by groups of strong Ly$\alpha$ absorption systems. The absorption group is defined as $\ge 4$ absorptions with $\tau^{15h^{-1}{\rm{Mpc}}}_{\rm{eff}}>3\times\mean{\tau_{\rm{eff}}}$ on 20 $h^{-1}$ cMpc, and each group contains at least a CoSLA candidate with $\tau^{15h^{-1}{\rm{Mpc}}}_{\rm{eff}}\ge 4.5\times\mean{\tau_{\rm{eff}}}$. Such absorption groups trace overdensities within the range of $\delta_m=0.6$ -- 2.8, with a median value of 1.50. From our LAE observations,  BOSS1441 ($\delta_m= 2.2$) is within this range.

\figurenum{8}
\begin{figure*}[tbp]
\epsscale{1.0}
\label{fig:02+04}
\plotone{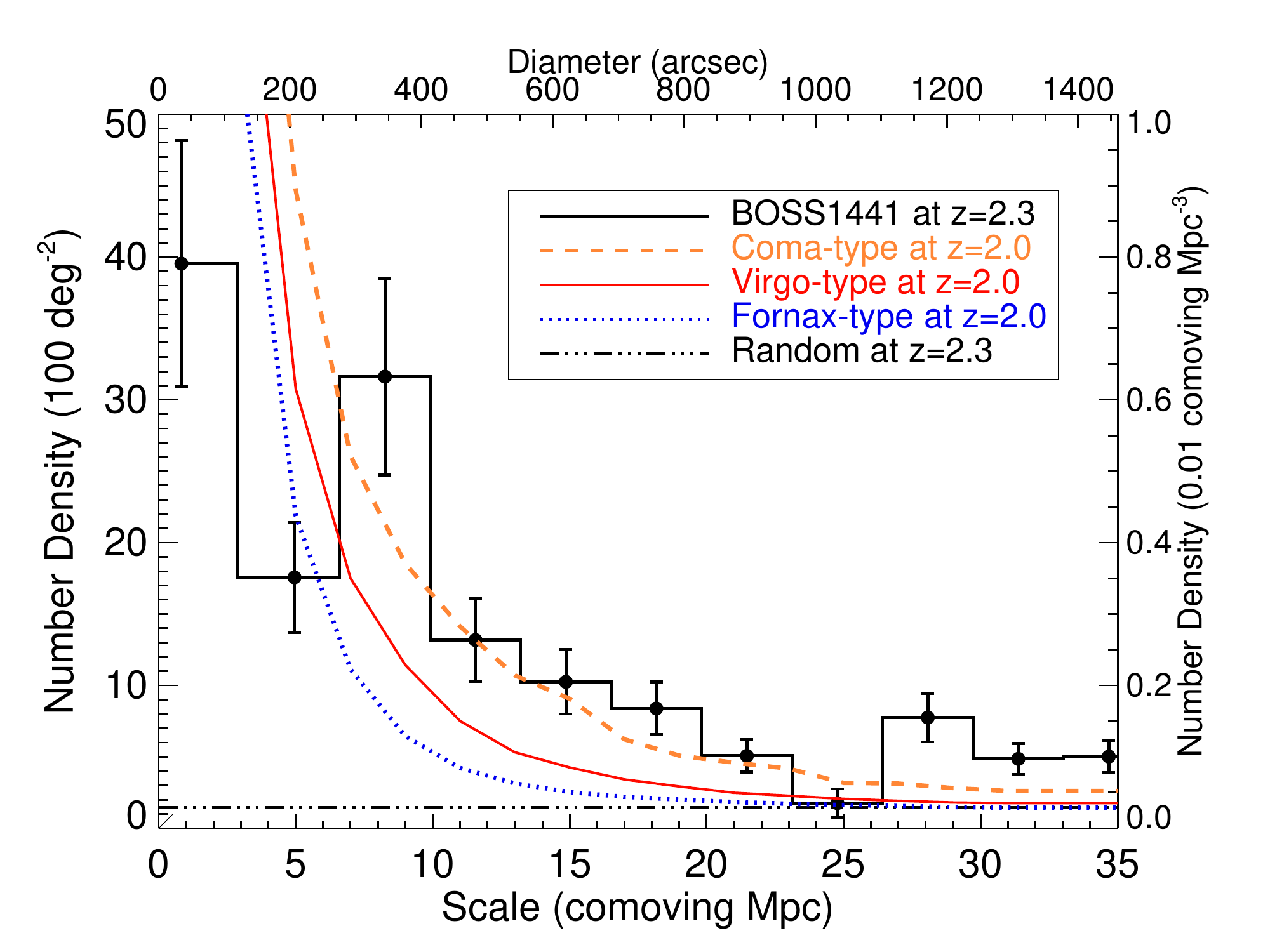}
\caption{The radial density profile measured in circular annuli centered on the density peak of BOSS1441  ($\alpha=14$:41:26.40,  $\delta=+40$:01:12.0).  The dotted horizontal line represents the number density at $z=2.3$ in random fields, calculated using the galaxy luminosity function in Ciardullo et al. (2012). This protocluster has one of the highest density of Ly$\alpha$ emitters found at $z=2-3$. We overplot the predicted radial density profile of different cluster progenitors at $z=2$ from  cosmological simulation (Chiang et al. 2013). The orange dashed line represents the progenitors of Coma-like clusters ($M_{\rm{z=0}}> 10^{15}\ M_\odot$), and the red represents the progenitors of Virgo-like clusters ($M_{\rm{z=0}}\approx 3-10\times10^{14}\ M_\odot$), and blue represents the progenitors of Fornax-like clusters  ($M_{\rm{z=0}}\approx 1-3\times 10^{14}\ M_\odot$).  }
\end{figure*}

\subsection{Rarity of this structure}

As shown in the previous section, BOSS1441 could collapse and evolve to a bound, virialized system at $z=0$ with a virialized mass ($M_{\rm{z\sim0}}\approx 10^{15}\ M_\odot$). 
Bahcall \& Cen (1993) present the cluster mass function at $z\lesssim0.2$. Extrapolating this cluster mass function  to the high-mass end (Bahcall et al. 2003; Casey 2016), we  estimate the volume density of high-mass clusters ($M>10^{15}\ M_\odot$)  is $\lesssim10^{-7}$, i.e., there is $\sim 1$ per (210 cMpc)$^3$ comoving volume. 
Using cosmological simulations, \citet{chiang13} also show that the abundance of protoclusters identified to have $M \gtrsim 10^{15}\ M_\odot$ at  $z\sim0$ is about 1--2 in a comoving box of $1\times10^7$ cMpc$^3$. Thus, our survey volume should contain $\sim 10$ protoclusters with $M \gtrsim 10^{15}\ h^{-1}\ M_\odot$. 

%These high-mass protoclusters all have $\delta_m > 2.2$ on 15 Mpc. If BOSS1441 collapse to a virilized cluster, then the abundance of BOSS1441 should 

%In order to understand how rare the BOSS1441 is, we need to understand our survey volume. 
 In the survey volume of $10^8$ cMpc$^3$ (see \S2.4), we indeed find 11 fields that contain the groups of CoSLAs. %But from Cai et al. (2015), these coherently strong absorption groups should trace a variety of mass overdensities. From the cosmological simulation, 30\% of the  them should have $\delta_m>2.2$ (Cai et al. 2015). 
% Every absorption group contains $\ge 4$ absorption systems in each independent sight lines with $\tau^{15h^{-1}{\rm{Mpc}}}_{\rm{eff}}> 3\times\mean{\tau_{\rm{eff}}}$ at $z=2.32\pm0.02$. BOSS1441 is one of these 11 fields.  
According to our cosmological simulation (Cai et al. 2015), the mass overdensity associated with the 11 absorption group has a wide distribution, with the range of $\delta_m=0.6$ -- 2.8. This wide range of mass overdensities arises from (1) the intrinsic scatter between $\tau_{\rm{eff}}$ and mass (see Figure~3 of Cai et al. 2015), and (2) 20\% of unidentified sub-DLA contaminants (see Cai et al. 2015). If the 11 fields have a mass distribution similar to our simulations, then 30\% of the  11 absorption ($\sim$ 3--4) groups should have $\delta_m>2.2$. 
Thus, although our first field BOSS1441 has an extremely high overdensity, the current BOSS background QSO density may not be high enough for completely selecting all the most massive overdensities in a survey volume.

\subsection{Redshift distribution of galaxies in BOSS1441}

The velocity dispersion measured from member galaxies is not a reliable mass measurement of a protocluster, because the protocluster is an unvirialized system. But some properties of protoclusters can still be reflected by the velocity dispersion. \citet{dey16} point out that the velocity dispersion could be affected by (a) the clumpiness of the protocluster region, (b) the dynamical evolution stage, and (c) the geometrical configuration between the protoclusters and the observed sight lines. % and (d) the narrowband or broadband selection technique. 
Using the Ly$\alpha$ emission, we calculate the redshifts for 20 galaxies (including one BOSS QSO) that are spectroscopically confirmed in the footprint of our MODS observations (50 arcmin$^2$).

Over the entire two LBT/MODS masks, the velocity dispersion is $\approx$863 km s$^{-1}$, with the median redshift of $z=2.319$. From higher  to lower redshifts, we list here a number of protoclusters that have velocity dispersion measurements over $\sim 5$ -- 15 cMpc scales. %A more comprehensive analysis can be found in Dey et al. (2016). 
Venemans et al. (2007) studied eight protoclusters traced by radio galaxies. %they measured that the velocity dispersions around radio galaxies range from $\sim$ 260 to 980 km s$^{-1}$. 
For the highest redshift protoclusters of TN J1338-1942 at  $z = 4.11$ and TN J0924-2201 at $z=5.20$, the velocity dispersions of $\approx 280$ km s$^{-1}$ and $\approx 305$ km s$^{-1}$, respectively. 
\citet{leeK14} discovered a large-scale structure at $z=3.78$ that \citet{dey16} show this has a relatively low velocity dispersion of ($\approx 350$ km s$^{-1}$) within 20 $h^{-1}$ cMpc. 
The SSA22 protocluster at $z=3.1$ has a velocity dispersion ranges from 400 -- 500 km s$^{-1}$ on each sub-group which extends over $\approx 15$ cMpc \citep{matsuda05, dey16}.  \citet{wang16} discovered an extremely overdense field at $z=2.51$ with a velocity dispersion of $\approx 500$ km s$^{-1}$ over core regions of 4 cMpc scale. Venemans et al. (2007) find that MRC 1138-262 at $z = 2.16$ has a high velocity dispersion of $\approx$ 900 km s$^{-1}$ on 10 cMpc scale,  which is similar to BOSS1441. %From all these measurement, it seems the velocity dispersion have a relatively higher value with 
%It is true that one should be cautious to derive conclusions, e.g., mass evolution, from the velocity dispersion with
%redshift, because the velocity dispersion measurements can be compromised by several reasons listed in the previous paragraph.
Compared with all the previously confirmed overdensities at $z=2-5$,  BOSS1441 has a relatively high velocity dispersion, suggesting that it may have the similar  dynamical maturity as the most massive overdensities  traced by radio galaxies at $z=2$. 

%In our next paper, we will further present our Keck/LRIS follow-up which obtains $>100$ galaxy spectra in BOSS1441 and completely reveal the galaxy properties in BOSS1441. 

\section{Summary and Future Prospects}

In this paper, we present an extremely massive overdensity BOSS1441 at $z=2.32\pm0.02$, discovered by utilizing our new approach of identifying overdensities from concentrations of coherently strong Ly$\alpha$ absorption systems (CoSLAs) \citep{cai15}.  BOSS1441 is traced by a group of coherently strong Ly$\alpha$ absorption with $\tau_{\rm{eff}}\ge 3.0\times\mean{\tau}$ within 15 $h^{-1}$ cMpc. The absorption group is defined as $\ge4$ absorption systems within the projected $20 h^{-1}$ cMpc scale and at $z=2.32\pm0.02$ (Figure~2 and Figure~3).  BOSS1441 is also associated with multiple BOSS QSOs at similar redshift. 

Our KPNO-4m/MOSAIC narrowband and multiple-wavelength broadband observations have revealed a large-scale structure extending over 25 $h^{-1}$ cMpc scale.  Using two LBT/MODS masks, we have spectroscopically confirmed 19 LAEs (Figure~7 and Table~3). The LAE density is $11.8\pm 2.6\times$ that in random fields in (15 cMpc)$^3$ volume ($\delta_g=10.8\pm2.6$), making BOSS1441 one of the most overdense fields discovered to date. 
We estimate that BOSS1441 should collapse to a virialized cluster with $M_{\rm{z=0}}\gtrsim 10^{15}$ M$_\odot$. The number of such massive clusters should be $\sim 1$ within a $10^7$ cMpc$^3$ volume. 

More observations are required to study the galaxy properties in such an overdense environment at $z=2.3$. We will use the {\it Hubble Space Telescope} Wide Field Camera 3 (WFC3) to conduct deep imaging in this field. These observations will test the existence of density-morphology relations  at $z=2.3$.  Also, we plan to enlarge our  sample of large-scale structures at $z=2-3$, to fully constrain the cosmic structure formation and better characterize the envioronmental dependence of galaxy properties. In near future, we aim to construct a uniform sample of the most massive overdensities at $z=2-3$ traced by CoSLAs. We have conducted a narrowband and broadband survey on other candidate fields by utilizing KPNO-4m/MOSAIC and CFHT/MagaCAM. Future multiple wavelength follow-ups and spectroscopic endeavors will quantify the overdensities, the diversity of galaxy populations, and the interactions between galaxy and IGM in the density peak at $z=2-3$. 

{{\bf Acknowledgement: }  ZC acknowlendges the insightful comments from Yi-Kuan Chiang, Arjun Dey, and Catlin Casey. ZC acknowledges Yuguang Chen's helps on the LBT/MODS observations.  ZC, XF, and IM thank the support from the US NSF grant AST 11-07682. ZC and JXP acknowledge support from NSF AST-1412981. AZ acknowledges support from NSF grant AST-0908280 and NASA grant ADP-NNX10AD47G. NK acknowledges supports from the JSPS grant 15H03645. Based on observations at Kitt Peak National Observatory, National Optical Astronomy Observatory (NOAO Prop. ID: 2013A-0434; PI: Z. Cai; NOAO Prop. ID: 2014A-0395; PI: Z. Cai), which is operated by the Association of Universities for Research in Astronomy (AURA) under cooperative agreement with the National Science Foundation. The authors are honored to be permitted to conduct astronomical research on Iolkam Du'ag (Kitt Peak), a mountain with particular significance to the Tohono O'odham. The LBT is an international collaboration among institutions in the United States, Italy and Germany. The LBT Corporation partners are: The University of Arizona on behalf of the Arizona university system; Istituto Nazionale di Astrofisica, Italy;  LBT Beteiligungsgesellschaft, Germany, representing the Max Planck Society, the Astrophysical Institute Potsdam, and Heidelberg University; The Ohio State University; The Research Corporation, on behalf of The University of Notre Dame, University of Minnesota and University of Virginia. }

\begin{table*} [!h]
\caption{Galaxy overdensity of BOSS1441} 
\label{table:PopIII_SFR}	
% is used to refer this table in the text
\centering 
\begin{tabular}{|c | c| c |c| c|c|c|} 
\hline\hline 
Volume & Depth & Random fields \footnotemark[1]  & MAMMOTH-1 & Noise & Over-& Mass within \\
  (cMpc)$^3$ & $L^*_{\rm{Ly\alpha}}$ & LAE   Number \footnotemark[2]   & LAE Number   &    number & denstiy & (15 cMpc)$^3$ (M$_\odot$)     \\  
% table heading
\hline
15$^3$ & 0.73 &  1.52 & 21 & 1.23 & $10.8\pm0.7$  &  $\sim1\times10^{15}$ \\
%\hline 
   % 23  &  55  & 25 $h^{-1}$ \\ 
%\hline 
 %     2  &   23 &  11 $h^{-1}$ (density peak including MAMMOTH-1) \\
\hline
\hline
\end{tabular}
\footnotetext[1]{Number calculated from Ly$\alpha$ luminosity function (Ciardullo et al. 2012) at $z=2.1$.  }
\footnotetext[2]{Number of Ly$\alpha$ emitters.  }
\end{table*}

\begin{table*} [!h]
\caption{Comparisons of  protoclusters tracing by different techniques at $z\sim2-3$}%($m\gtrsim5\times10^{14}$ M$_\odot$) 
\label{table:PopIII_SFR}	
% is used to refer this table in the text
\centering 
\begin{tabular}{|c | c| c |c |c|c|} 
\hline\hline 
Name  &  $\mean{z}$ &  Scale   & $\delta_g$ & Mass & Reference\\ 
\hline
    &                    &   (cMpc)       &              &  ($10^{14}$ M$_\odot$) &  \\
\hline
CDFSz28    &   2.82  &  15 &  [4.7, 4.7, 6.0, 6.6] &  [3.3, 3.3, 3.7, 3.8]  & 9 \\ 
COSMOS34 $^a$ & $3.04$ & 15 &  $2.28^{+1.04}_{-0.79}$ &   ... & 8 \\ 
 HS 1700+643  & 2.300 &  19.5  & 5.7 (6.9 in paper) & 12 & 2,3 \\ 
J1431+3239 &  $3.733$ & $ 15$ &  $4.2\pm0.9$ & $7$ & $5$ \\
MRC0943-242   & 2.92  &   20  &  $2.2^{+0.9}_{-0.7}$ & 4-5 & 4,5\\
Slug $^b$   &    2.28 &  10 &   modest &  ...  &  10 \\
 SSA22 &      3.09 &   23 &   $3.6^{+1.4}_{-1.2}$  &  $\sim 10$ &  1     \\
 TN J1338-1942$^c$ &  $4.1$ &  $ 18.$ &  3.8$^{+1.1}_{-0.8}$ & 7 & 4,5,6 \\
This work &  $2.329\pm0.013$ &  15 &  $10.8\pm2.6$ & $10$ &     \\
This work &   $2.329\pm0.013$   &  19.5  &   $8.3\pm 2.0 $  &  20 & \\
% table heading
\hline
\hline
\end{tabular}
\\
       {{\bf{Note: }} Summary of massive overdensities discovered at $z\sim2-3$ with mass $>5\times10^{14}$ M$_\odot$. The mass is determined using our consistent way by setting galaxy bias $b=2.1$. SSA22, COSMOS34, J1431+3239 and HS1700+643 are three overdensities discovered using galaxy redshift survey.

a: COSMOS34 has the highest $\delta_g$ in all protoclusters in cosmoc field (Chiang et al. 2013). 

b: TNJ1338-1942 is the most massive protoclusters discovered using radio-galaxies as a tracer (Venemans et al. 2007). 

c: Slug is the largest Ly$\alpha$ nebula that have previous discovered before this work (Cantalupo et al. 2014). 
{\bf References}: (1) Steidel et al. 1998; (2) Steidel et al. 2005; (3) Digby-North et al. 2010; (4) Venemans et al. 2007; (5) Venemans et al. 2005;  (6) Venemans et al. 2002; (7) Lee et al. 2014; (8) Chiang et al. 2014; (9) Zheng et al. 2016; (10) Cantalupo et al. 2014; }

%\footnotetext[1]{Steidel et al. (1998) , Steidel et al. (2005),  Digby-North et al. (2010)}
\end{table*}

\begin{table*} [!h]
\caption{19 Spectroscopic confirmed galaxies at $z=2.32\pm0.02$ in our MODS follow-up observations (50 arcmin)$^2$. } 
\label{table:spec}	
% is used to refer this table in the text
\centering 
\begin{tabular}{ c| c | c |c | c } 
\hline\hline 
Name & RA & DEC & $L_{\rm{Ly\alpha}}$ (erg s$^{-1}$)& redshift \\
\hline
LAE1 & 14:41:44.67   & +40:00:13.52 & $3.75\pm0.2\times10^{42}$ & 2.312 \\
\hline
LAE2 & 14:41:43.58    & +40:03:31.15 &  $8.40\pm0.2\times10^{42}$   & 2.313  \\
\hline
LAE3 & 14:41:40.86   & +40:03:05.21 &  $4.60\pm0.2\times10^{42}$   &  2.323 \\
\hline
LAE4 & 14:41:39.97  & +40:02:30.07& $2.38\pm0.2\times10^{42}$   &  2.330  \\
\hline 
LAE5 & 14:41:38.04   & +40:01:08.39 &   $15.1\pm0.2\times10^{42}$  & 2.338  \\
\hline
LAE6 & 14:41:31.01 & +40:01:09.72 &  $2.34\pm0.2\times10^{42}$    & 2.321 \\
\hline
LAE7 & 14:41:27.98 &  +40:01:10.79  & $3.53\pm0.2\times10^{42}$  & 2.312  \\
\hline     
LAE8 & 14:41:26.22 & +40:04:12.02 &   $3.35\pm0.2\times10^{42}$ & 2.312   \\
\hline
LAE9 & 14:41:25.30  & +40:03:55.89 &  $8.9\pm0.2\times10^{42}$   &  2.319 \\
\hline 
LAE10 (MAMMOTH-1 Nebula) & 14:41:24.31 &  +40:03:11.91  &   $528\pm0.3\times10^{42}$  & 2.319   \\
\hline 
 LAE11 & 14:41:33.82  & +40:01:42.75 &     $3.62\pm0.3\times10^{43}$  & 2.311   \\
\hline
LAE12 & 14:41:20.39  & +39:53:55.39 &   $14.4\pm0.3 \times10^{43}$ & 2.336 \\
\hline
LAE13 & 14:41:15.03 &  +39:54:12.10 & $7.3\pm0.3\times10^{42}$ & 2.307  \\ 
\hline
LAE14 & 14:41:14.15 & +39:55:35.57 &  $3.9\pm0.3\times10^{42}$ & 2.334 \\ 
\hline
LAE 15 (AGN) & 14:41:11.72   &  +39:56:54.96 &  $234.6\pm0.7\times10^{42}$ & 2.311 \\ 
\hline
LAE16 & 14:41:08.10 &  +39:55:02.62 &   $3.2\pm0.3 \times10^{42}$ & 2.309 \\
\hline
LAE17 & 14:41:04.83  &  +39:54:02.31 &   $3.5\pm0.3 \times10^{42}$ & 2.326 \\ 
\hline 
LAE18 & 14:41:01.51 & +39:55:31.41  &   $4.2\pm0.3 \times 10^{42}$ & 2.317 \\ 
\hline
LAE19 & 14:40:54.20  &  +39:52:51.62 &   $1.7\times0.3\times 10^{42}$ & 2.308 \\
\hline 
\hline
\end{tabular}
\end{table*}

\begin{table*} [!h]
\caption{BOSS QSOs in BOSS1441} 
\label{table:spec}	
% is used to refer this table in the text
\centering 
\begin{tabular}{ c| c | c | c } 
\hline\hline 
Name & RA & DEC  & redshift \\
\hline
BG QSO1$^a$ & 14:40:48.56   &   +39:56:18.3  & 2.542 \\
\hline
BG QSO2 & 14:39:58.32   & +40:03:14.0   & 2.422 \\
\hline
BG QSO3 & 14:41:10.98   &   +39:54:23.9 & 2.999 \\
\hline 
BG QSO4 & 14:42:51.83   & +40:14:53.4  & 2.553 \\
\hline
BG QSO5 & 14:42:10.59   &   +39:56:31.8 & 2.612 \\
\hline 
BG QSO6 &  14:39:58.44   &  +40:03:14.0  &  2.425 \\
\hline
QSO1 & 14:40:27.91    &   +40:11:47.8  & 2.324   \\
\hline
QSO2 & 14:40:49.15     &  +39:54:07.6  & 2.323   \\
\hline
QSO3$^b$ &  14:41:21.67    & +40:02:58.9  &  2.311  \\
\hline
QSO4 & 14:40:57.39     & +39:49:51.2  &  2.324   \\
\hline 
QSO5 & 14:41:15.91    &  +39:49:12.0  & 2.334   \\
\hline 
QSO6 &  14:41:32.26   &   +40:06:35.6  & 2.311  \\
\hline
\end{tabular}
\footnotetext[1]{All the background QSOs in this field have strong Ly$\alpha$ absorption detected at $z=2.32\pm0.02$ (Figure~2, Figure~3).}
\footnotetext[2]{This QSO resides in the (15 cMpc)$^3$ density peak region of BOSS1441. }
\end{table*}

%\newpage

\end{document}